\documentclass[showpacs,preprintnumbers,twocolumn,amsmath,amssymb,10pt]{revtex4}

\usepackage{graphicx}
\usepackage{dcolumn}
\usepackage{bm}
\usepackage{hyperref}
\usepackage{color}
\usepackage{paralist}

\definecolor{Blue}{rgb}{0.3,0.3,0.9}
\definecolor{Red}{rgb}{1,0,0}
\definecolor{Green}{rgb}{0,1,0}

\begin{document}
\title{Inflation in loop quantum cosmology: \\ Dynamics and spectrum of gravitational waves }

\author{Jakub Mielczarek}
\email{jakub.mielczarek@uj.edu.pl}
\affiliation{Astronomical Observatory, Jagiellonian University, 30-244
Krak\'ow, Orla 171, Poland}

\author{Thomas Cailleteau}
\email{cailleteau@lpsc.in2p3.fr}
\affiliation{
Laboratoire de Physique Subatomique et de Cosmologie, UJF, INPG, CNRS, IN2P3\\
53, av. des Martyrs, 38026 Grenoble cedex, France
}

\author{Julien Grain}
\email{julien.grain@ias.u-psud.fr}
\affiliation{
Institut d'Astrophysique Spatiale, Universit\'e Paris-Sud 11, CNRS \\ B\^atiments 120-121, 91405 Orsay Cedex, France }

\author{Aurelien Barrau}
\email{aurelien.barrau@cern.ch}
\affiliation{
Laboratoire de Physique Subatomique et de Cosmologie, UJF, INPG, CNRS, IN2P3\\
53, avenue des Martyrs, 38026 Grenoble cedex, France
}

\date{\today}

\begin{abstract}

Loop quantum cosmology provides an efficient framework to study the evolution of the Universe
beyond the classical Big Bang paradigm. Because of holonomy corrections, the singularity is
replaced by a ``bounce."  The dynamics of the background is investigated into the details,
as a function of the parameters of the model. In particular, the conditions
required for inflation to occur are carefully considered and are shown to be generically met. 
The propagation of gravitational waves is then investigated in this framework. By both numerical and
analytical approaches, the primordial tensor power spectrum is computed for a wide range of
parameters. Several interesting features could be observationally probed.

\end{abstract}

\pacs{04.60.Pp, 04.60.Bc, 98.80.Cq, 98.80.Qc}
\keywords{Quantum gravity, quantum cosmology, inflation, gravitational waves}

\maketitle

\section{Introduction}

Loop quantum gravity (LQG) is a nonperturbative and background-independent quantization
of general relativity. Based on a canonical approach, it uses Ashtekar variables, namely
SU(2) valued connections and conjugate densitized triads. The quantization is 
obtained through holonomies of the connections and fluxes of the densitized triads (see, 
{\it e.g.}, \cite{rovelli1} for an introduction). Basically, loop quantum cosmology (LQC)
is the symmetry reduced version of LQG (although it is fair to underline that the relations
with the full theory are still to be investigated into the details). While predictions of LQC
are very close to those of the old quantum geometrodynamics theory in the low curvature
regime, there is a dramatic difference once the density approaches the Planck scale: the big bang
is replaced by a big bounce due to huge repulsive quantum geometrical effects 
(see, {\it e.g.}, \cite{lqc_review} for a review).
Among the successes of LQC, one can cite: the excellent agreement between the trajectories
obtained in the full quantum theory and the classical Friedman dynamics as far as the
density in much below the Planck scale, the resolution of past and future singularities, 
the ``stability" of states which remain sharply peaked even after many cycles 
(in the k=1 case) and the fact that initial conditions for inflation are somehow 
naturally met. The latter point is especially appealing as the inflationary scenario is 
currently the
favored paradigm to describe the first stages of the evolution of the Universe
(see, {\it e.g.}, \cite{linde} for a recent review). Although still debated, 
it has received many experimental confirmations, including from the
WMAP 7-Years results \cite{wmap}, and solves most cosmological paradoxes. It is rather
remarkable that, as will be explained in this paper, the canonical quantization of general
relativity naturally leads to inflation without any fine tuning. Inflation would have
been unavoidably predicted by LQC, independently of its usefulness in the cosmological
paradigm.\\

Two main quantum corrections are expected from the Hamiltonian of LQG when dealing with a
semiclassical approach, as will be the case in this study mostly devoted to potentially
observable effects. The first one comes
from the fact that loop quantization is based on holonomies, {\it i.e.} exponentials of the
connection rather than direct connection components. The second one arises for inverse
powers of the densitized triad, which when quantized become an operator with 
zero in its discrete spectrum thus lacking a direct inverse.  
As the status of "inverse volume" corrections is not clear due to 
the fiducial volume cell dependence, this work focuses on the holonomy term
only and derives, for the first time in a fully consistent way, 
the entire dynamics up to the explicit computation of the tensor power
spectrum. The background evolution is first studied and a specific attention is paid to the
investigation of the inflationary stage following the bounce. Then, analytical formulas
are given for the primordial tensor spectrum for either a pure de Sitter or a slow-roll
inflation. Finally, numerical results are given for many values of the parameters of the
model.

\section{Background dynamics} \label{SecII}

In general, many different evolutionary 
scenarios are possible within the framework of LQC.
However, all of them have a fundamental common feature, namely 
the cosmic bounce. As we will show, the implementation of a suitable matter content also 
generically leads to a phase of inflation. This phase is nearly mandatory in any meaningful cosmological
scenario since our current understanding of the growth of cosmic structures 
requires --among many other things-- inflation in the early universe. It is therefore important
to study the links between the inflationary paradigm and the LQC framework, as emphasized, {\it e.g.},
in \cite{Ashtekar:2009mm}.

The demonstration that a phase of superinflation can occur due to 
quantum gravity effects was one of the first great achievements 
of LQC \cite{bojo2002}. This result was  based on the so-called inverse volume corrections.
It has however been understood that such 
corrections exhibit a fiducial cell dependence,  making the 
physical meaning of the associated results harder to understand.
As reminded in the introduction, other corrections also arise in LQC, due to so-called holonomy 
terms, which do not depend on the fiducial cell volume. Those corrections lead to  
a dramatic modification of the Friedmann equation which becomes
\begin{equation}
H^2 =\frac{\kappa}{3} \rho \left(1-\frac{\rho}{\rho_{\text{c}}}  \right), \label{Friedmann}
\end{equation}
where $\rho$ is the energy density, $\rho_{\text{c}}$ is the critical energy density,  
$H$ is the Hubble parameter, and $\kappa=8\pi G$. In principle, $\rho_{\text{c}}$ can be viewed as a 
free parameter of theory. However, its value is usually determined 
thanks to the results of area quantization in LQG. Then, 
\begin{equation}
\rho_{\text{c}} = \frac{\sqrt{3}}{16\pi^2\gamma^3}m^4_{\text{Pl}} \simeq 0.82 m^4_{\text{Pl}},  \label{rhoc}
\end{equation}  
where value $\gamma \simeq 0.239$ has been used, as obtained from the computation
of the entropy of black holes \cite{Meissner:2004ju}. Should the inverse volume corrections
be included, this would modify the background dynamics by some additional factors.

\begin{figure}[ht!]
\centering
\includegraphics[width=8cm,angle=0]{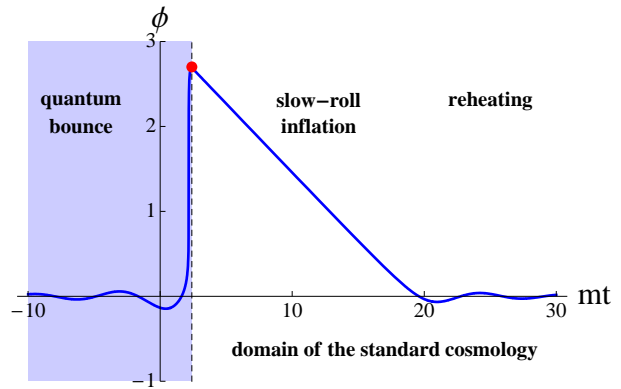}
\caption{\emph{Shark fin}-type evolution of a scalar field for 
$m=10^{-3}m_{\text{Pl}}$. The (red) dot represents the point where the 
initial conditions in  classical cosmology are usually set.}
\label{field1}
\end{figure}

As it can easily be seen from Eq.~(\ref{Friedmann}), a general prediction associated
with models including holonomy corrections is a bounce which
occurs for  $\rho=\rho_{\text{c}}$. The appearance of this $\rho^2$ term with the correct
negative sign is a highly nontrivial and appealing feature of this framework which
shows that the repulsive quantum geometrical effects become dominant in the Planck region.
The very quantum nature of spacetime is capable of overwhelming the huge gravitational attraction.
The dynamics of models with holonomy corrections
was studied in several articles \cite{Ashtekar:2009mm,Singh:2006im,Mielczarek:2009zw,Chiou:2010nd}. 
In this paper we further perform a detailed and consistent study 
of a universe filled with a massive scalar field in this framework. 
The global dynamics of such models was firstly studied in Ref.~\cite{Singh:2006im}.
Recently, it was pointed out in Ref.~\cite{Mielczarek:2009zw}  that
the "standard" slow-roll inflation is triggered by the preceding phase of 
quantum bounce. This general effect is due to the fact that the universe
undergoes contraction before the bounce, resulting in a  negative 
value of the Hubble factor $H$. Since the equation governing the evolution 
of a scalar field in a Friedmann-Robertson-Walker universe is 
\begin{equation}
\ddot{\phi}+3H\dot{\phi}+m^2\phi=0, 
\end{equation}
the negative value of $H$ during the prebounce phase acts as an {\it antifriction} term leading to 
the amplification of the oscillations of field $\phi$. In particular, 
when the scalar field is initially at the bottom of the potential well
with some small nonvanishing derivative $\dot{\phi}$, then it is driven up the 
potential well as a result of the contraction of the universe. This situation
is presented in Fig.~\ref{field1}

To some extent, it is therefore reasonable to say that the LQC framework solves both the two main "problems"
of the big bang theory: the singularity (which is regularized and replaced by a bounce) and the initial
conditions for inflation (which are naturally set by the antifriction term). 

However, this \emph{shark fin} evolution (see caption of Fig.~\ref{field1}) is not the only possible one.
In particular, a nearly \emph{symmetric} evolution can also take place, 
as studied in Ref.~\cite{Chiou:2010nd}.
Those different scenarios can be distinguished by the fraction of kinetic 
energy at the bounce. When the energy density at the bounce 
is purely kinetic, the evolution of the field is symmetric.
When a small fraction of potential energy is introduced, which is the general case,
the symmetry is broken and the field behaves as in the \emph{shark fin} case. 
It is however important to underline that we consider only scenarios where the
contribution from the potential is subdominant at the bounce, as it would otherwise be necessary
to include quantum backreaction effects \cite{bojo:back}. The effective dynamics
would then be more complicated and could not be anymore described by Eq.~(\ref{Friedmann}).

In order to perform qualitative studies of the dynamics of the model, it is  useful 
to introduce the variables
\begin{equation}
x := \frac{m\phi}{\sqrt{2\rho_{\text{c}}}} \  \text{and} \  y :=\frac{\dot{\phi}}{\sqrt{2\rho_{\text{c}}}} . \label{xy}
\end{equation}
Since the energy density of the field is constrained ($\rho \leq \rho_{\text{c}}$), the inequality
\begin{equation}
x^2+y^2 \leq 1
\end{equation}
has to be fulfilled.
The $x^2$ term corresponds to the potential part while the $y^2$ corresponds to the kinetic term.
The case $x^2+y^2 =1$ corresponds to the bounce, when the energy density reaches its maximum.

\begin{figure}[ht!]
\centering
\includegraphics[width=7cm,angle=0]{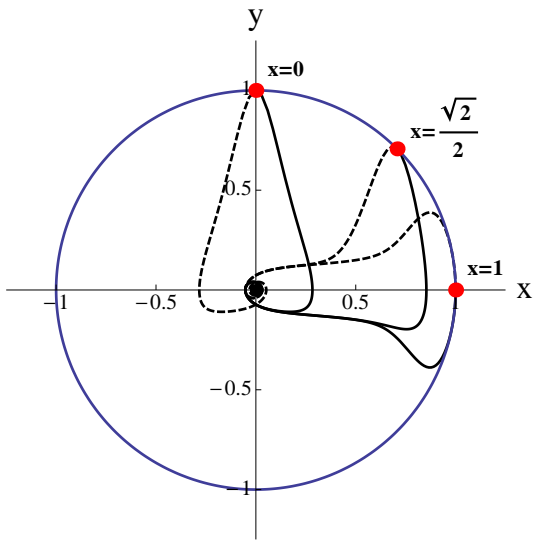}
\caption{Exemplary phase trajectories of the scalar field with $m=m_{\text{Pl}}$.}
\label{PP}
\end{figure}

In Fig.~\ref{PP}, exemplary evolutionary paths in the $x-y$ phase plane are shown. 
For all the presented cases, the evolution begins at the origin (in the limit
$t\rightarrow -\infty$), and then evolves (dashed line)
to the point on the circle $x^2+y^2=1$. Finally, the field moves back to the origin 
for $t\rightarrow +\infty$ (solid line). However, the shapes of the intermediate paths are different.  
The $x=0$ case corresponds to the \emph{symmetric} evolution which was studied
in Ref.~\cite{Chiou:2010nd}  (if the bounce is set at $t=0$, the scale factor is an even function of
time and the scalar field is an odd function). In this case, the field is at the bottom of the potential well
exactly at the bounce ($H=0$). This is however a very special choice of initial conditions.  
In the case $x=\sqrt{2}/2$, the potential term and kinetic term contribute equally at the 
bounce. In this case, both deflation and inflation occur. However one observes 
differences in their duration. The third case, $x=1$, corresponds to the domination of the 
potential part at the bounce. In this case, symmetric phases of deflation and inflation 
also occur (both the scale factor and the field being this time even functions).
However in this situation, as well as in $x=\sqrt{2}/2$ case, the effect of 
quantum backreaction should be taken into account. The dynamics can therefore significantly 
differ from the one computed with Eq.~(\ref{Friedmann}). 

\begin{figure}[ht!]
\centering
\includegraphics[width=8cm,angle=0]{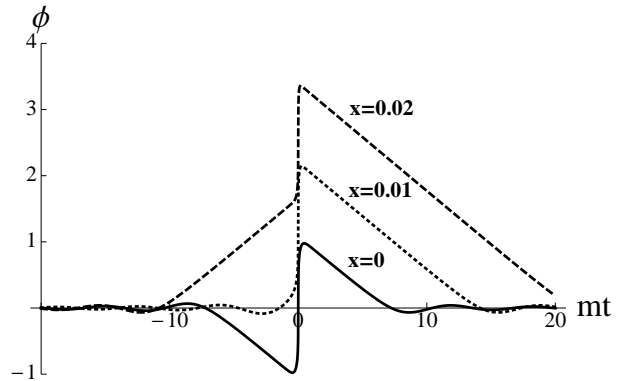}
\caption{Time evolution of the scalar field. Different evolutionary scenarios leading to 
a slow-roll inflation phase are displayed. The bottom (solid) line represents the 
\emph{symmetric} case. 
The middle (dotted) line represents the \emph{shark fin}-type evolution mostly investigated in this paper. 
The top (dashed) line corresponds to a larger fraction of potential energy. For all curves 
$m=0.01 m_{\text{Pl}}$.}
\label{field2}
\end{figure}

\begin{figure}[ht!]
\centering
\includegraphics[width=8cm,angle=0]{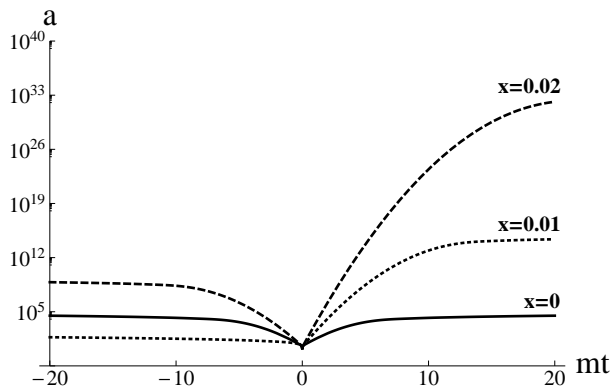}
\caption{Time evolution of the scale factor. Different evolutionary scenarios leading 
to a slow-roll inflation phase are displayed. The bottom (solid) line represents the \emph{symmetric} case. 
The middle (dotted) line represents the \emph{shark fin} type evolution. 
The top (dashed) line corresponds to a larger fraction of potential energy. For all curves 
 $m=0.01 m_{\text{Pl}}$.}
\label{scalefactor}
\end{figure}

In Fig.~\ref{field2} we show some exemplary evolutions of the scalar field for
different contributions from the potential part at the bounce.
In Fig.~\ref{scalefactor}, the corresponding evolutions of the scale factor are displayed.
It can easily be seen that the value of $\phi_{\text{max}}$ increases with the fraction of 
potential energy 
at the bounce. Since the total energy density is constrained, $\phi_{\text{max}}$ must satisfy
\begin{equation}
|\phi_{\text{max}}| \leq \frac{\sqrt{2\rho_{\text{c}}}}{m}. \label{phimax}
\end{equation}
The values of $\phi_{\text{max}}$ associated with different evolutionary scenarios were computed
in \cite{Ashtekar:2009mm,Mielczarek:2009zw,Chiou:2010nd}. The conclusion of those
studies is that the necessary conditions for inflation are generically met. Only in the case of a \emph{symmetric}
evolution does the value of $\phi_{\text{max}}$ become too small in some cases.
In particular, for $m=10^{-6} m_{\text{Pl}}$ one obtains $\phi_{\text{max}}=2.1m_{\text{Pl}}$  
for a \emph{symmetric} evolution. 
The corresponding number of e-folds can be computed with
$N\simeq 2\pi \frac{\phi^2}{m^2_{\text{Pl}}}$, which gives $N\simeq 28$.
By introducing a small fraction of potential energy (as in the \emph{shark fin} case), the number
of  e-folds can be appropriately increased. In addition to the usual arguments, this requirement is also
set by the recent WMAP 7-Years results \cite{wmap}. Based on those observations, the value of the 
scalar spectral index was indeed measured to be $n_{\text{S}}=0.963\pm0.012$. As, for a 
massive slow-roll inflation the relation
\begin{equation}
n_{\text{S}} = 1-\frac{1}{\pi}\frac{m^2_{\text{Pl}} }{\phi^2}
\end{equation}
holds, one obtains $\phi_{\text{obs}}=2.9\pm0.5 m_{\text{Pl}}$.  
Since the consistency relation $\phi_{\text{max}}>\phi_{\text{obs}}$ must
be fulfilled, the symmetric evolution with $m=10^{-6} m_{\text{Pl}}$ 
(for which $\phi_{\text{max}}=2.1m_{\text{Pl}}<\phi_{\text{obs}}$) is 
not favored by the WMAP 7-Years observations. As  already mentioned, 
higher values of $\phi$ can be easily reached if some contribution from the 
potential term is introduced (this supports the \emph{shark fin} scenario).
The number of e-folds will therefore be naturally increased in this way.
However it remains bounded by above:  since $N\simeq 2\pi \frac{\phi^2}{m^2_{\text{Pl}}}$, 
Eq.~(\ref{phimax}) leads to the constrain:
\begin{equation}
N \leq \frac{4\pi \rho_{\text{c}}}{m^2m^2_{\text{Pl}}}.
\end{equation}

\begin{figure}[ht!]
\centering
\includegraphics[width=8cm,angle=0]{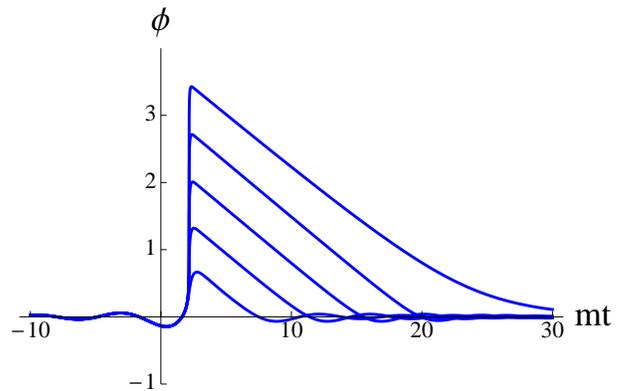}
\caption{The \emph{shark fin}-type evolution of the scalar field for 
$m=10^{-3}m_{\text{Pl}}$. Curves from bottom to top were computed for 
$\rho_{\text{c}}= 10^{-6}, 10^{-4}, 10^{-2}, 1, $ and $100 \ [m^4_{\text{Pl}}],$ respectively.}
\label{field3}
\end{figure}

The value of the parameter $\rho_{\text{c}}$ can be fixed by Eq.~(\ref{rhoc}).
However, this expression is based on the computation of the area gap as performed in
LQG. This, in general, can be questioned \cite{Dzierzak:2008dy}. In particular, 
in the framework of reduced phase space quantization of LQC, the value of 
$\rho_{\text{c}}$ remains a free parameter \cite{Malkiewicz:2009zd}. Moreover, 
a particular value of the Barbero-Immirzi parameter 
(imposed by black hole entropy considerations) has been used. Therefore, the 
value of $\rho_{\text{c}}$ can, in general, differ and it is worth investigating 
how the variation of $\rho_{\text{c}}$ can alter the dynamics of the model. In 
particular, we have studied how the \emph{shark fin} scenario can be modified by 
different choices of $\rho_{\text{c}}$.
In Fig.~\ref{field3}, the evolution of the field is displayed as a function of the value of the
critical energy density. As expected, the larger $\rho_{\text{c}}$, the higher the maximum value reached
by the field. It can be seen that $\phi_{\text{max}}$ approaches the usually required value $\sim 3
m_{\text{Pl}}$ for $\rho_{\text{c}}\sim m^4_{\text{Pl}}$, making the whole scenario
quite natural.

\section{Gravitational waves in LQC}

Although quite a lot of work has already been devoted to gravitational waves in 
LQC \cite{lqcgen}, this study aims at treating, for the first time, the problem in a fully self-consistent
way with an explicit emphasis on the investigation of the spectrum that can be used as an
input to study possible experimental signatures.

The equation for tensor modes in LQC is given (see, {\it e.g.}, \cite{Bojowald:2007cd}) by 
\begin{equation}
\frac{d^2}{d\eta^2}h^i_a + 2aH\frac{d}{d\eta}h^i_a-\nabla^2h^i_a+
m^2_Qh^i_a = 0,
\end{equation}
where $h^i_a$ are gravitational perturbations, $\eta$ is the conformal time and the factor
due to the holonomy corrections is given by  
\begin{equation}
m^2_Q := 16 \pi G a^2 \frac{\rho}{\rho_c}\left(\frac{2}{3}\rho-V \right).       
\end{equation}
This factor acts as an effective mass  term. For  convenience we introduce the variable
\begin{equation}
u=\frac{a h_{\oplus}}{\sqrt{16\pi G}}=\frac{a h_{\otimes}}{\sqrt{16\pi G}},  
\end{equation}
where $h^1_1=-h^2_2=h_{\oplus}$, $h^1_2=h^2_1=h_{\otimes}$.
Then, performing the Fourier transform
\begin{equation}
u({\bf x},\eta) = \int \frac{d^3{\bf k}}{(2\pi)^3}  u_{\bf k}(\eta) e^{i{\bf k}\cdot {\bf x}}, 
\end{equation}
one can rewrite the equation as
\begin{equation}
\frac{d^2}{d\eta^2} u_{\bf k}(\eta)+\left[k^2+m^2_{\text{eff}}\right]u_{\bf k}(\eta)=0, \label{ModeEQ}
\end{equation}
where $k^2={\bf k}\cdot {\bf k}$ and 
\begin{equation}
m^2_{\text{eff}}  := m^2_Q - \frac{a^{''}}{a}=a^2 \frac{\kappa}{2}\left[p-\frac{1}{3}\rho\right]. \label{meffdef}
\end{equation}
It is worth underlining that the final expression of $m_{\text{eff}}$ has no explicit  
dependence upon the 
critical energy density $\rho_{\text{c}}$. In Eq.~(\ref{meffdef}), both $m^2_Q$ and $a^{''}/a$ 
depend on $\rho_{\text{c}}$. However since 
\begin{equation}
\frac{a^{''}}{a} = 
a^2\left[\frac{2\kappa}{3} \rho \left(1-\frac{\rho}{\rho_{\text{c}}}\right)-
\frac{\kappa}{2}(\rho+p)\left(1-\frac{2\rho}{\rho_{\text{c}}}\right) \right],
\end{equation}
the factors depending on $\rho_{\text{c}}$ cancel out precisely. This is perhaps 
not a coincidence and this could exhibit the conservation of classical symmetries 
while introducing the quantum corrections.  
  
The next step consists in quantizing the Fourier modes $u_{\bf k}(\eta)$. 
This follows the standard canonical procedure. Promoting this quantity to be an
operator, one performs the decomposition
\begin{equation}
\hat{u}_{\bf k}(\eta) = f_k(\eta) \hat{b}_{\bf k} +f_k^*(\eta) \hat{b}^{\dagger}_{\bf-k}, \label{decomp1}
\end{equation}
where $f_k(\eta)$ is the so-called  mode function which satisfies the same equation as 
$u_{\bf k}(\eta)$, namely Eq.~(\ref{ModeEQ}). The creation ($\hat{b}^{\dagger}_{\bf k}$) 
and annihilation ($\hat{b}_{\bf k}$) operators fulfill the commutation relation $[\hat{b}_{\bf k},\hat{b}^{\dagger}_{\bf q}]
=\delta^{(3)}({\bf k-q})$. 
  
The problem is now shifted to the resolution of a Schr\"odinger-like  
Eq.~(\ref{ModeEQ}) which can be used to compute the observationally relevant 
quantities. In particular, the correlation function for tensor 
modes is given by 
\begin{eqnarray}
\langle 0|  \hat{h}^a_b({\bold x},\eta) \hat{h}^b_a ({\bold y},\eta)| 0 \rangle 
=\int_0^{\infty} \frac{dk}{k} \mathcal{P}_{\text{T}}(k,\eta) \frac{\sin kr}{kr},  
\end{eqnarray}
where $\mathcal{P}_{\text{T}}$ is the tensor power spectrum 
and $| 0 \rangle $ is the vacuum state.
In our case,  $\mathcal{P}_{\text{T}}$ can be written as
\begin{equation}
\mathcal{P}_{\text{T}}(k,\eta) = \frac{64 \pi G}{a^2 (\eta)} \frac{k^3}{2 \pi^2} |f_k(\eta)|^2.
\end{equation}
This spectrum is the fundamental observable associated with gravitational wave production. As will be
shown in the next sections, very substantial deviations from the usual shape are to be expected within
the LQC framework. 

\section{Analytical investigation of the power spectrum}

In this section we perform analytical studies of gravitational wave creation 
in the scenario previously described. In particular, we derive approximate formulas 
for the tensor power spectrum at the end of inflation. In the next section we will
compare this result with  numerical computations.
 
In the considered model, the evolution is split into three parts: contraction, bounce and 
slow-roll inflation. For this model, the effective mass square is defined as follows
\begin{equation}
m^2_{\text{eff}}(\eta) = \left\{ \begin{array}{ccc} 
0 & \text{for} &\eta < \eta_i-\Delta\eta.   \\ 
k_0^2  & \text{for} & \eta_i-\Delta\eta  < \eta <  \eta_i. \\
-\left(\nu^2-\frac{1}{4}\right) \frac{1}{\eta^2}  & \text{for} & \eta > \eta_i.   \end{array}  \right. 
\label{ameff}
\end{equation}
Basically, the phenomenological parameters entering the model are therefore:
\begin{itemize}
\item $\eta_i$ --- the beginning of the inflation.
\item $\Delta \eta$ --- the width of the bounce.
\item $k_0$ --- which is approximately equal to the value of $m_{\text{eff}}$ at the bounce 
(when $H=0$). It can therefore be related with the energy scale of the bounce.
\item $\nu$ --- which is related to slow-roll parameter $\epsilon$ by 
$\nu = \sqrt{\frac{9}{4}+3\epsilon}=\frac{3}{2}+\epsilon+
\mathcal{O}(\epsilon^2)$, where $\epsilon \ll 1$.
\end{itemize} 

For the considered model, we have $k^2_0 \geq 0$.  This comes from the fact that we
consider the particular \emph{shark fin}-type of evolution where the bounce is dominated by the 
kinetic energy term. 
Therefore when $y\gg x$ [see Eq.~(\ref{xy})], Eq.~(\ref{meffdef}) simplifies to  
$m^2_{\text{eff}} =a^2\kappa\dot{\phi}^2/6 \geq 0$, leading to 
$k^2_0\approx m^2_{\text{eff}}(t=t_{\text{bounce}})\geq 0$.

\begin{figure}[ht!]
\centering
\includegraphics[width=8cm,angle=0]{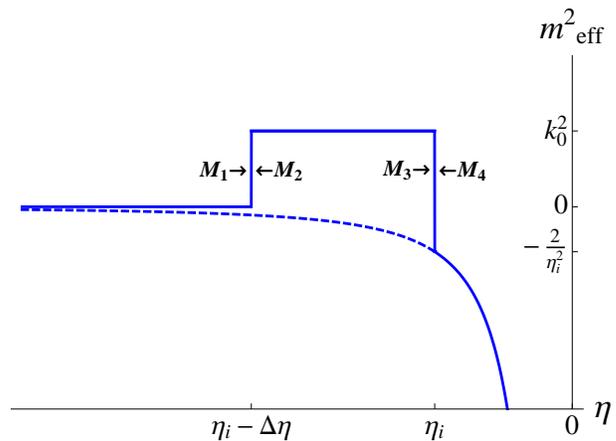}
\caption{Evolution of the effective mass used in the analytical approximation [Eq.~(\ref{ameff})]. 
On this plot, $\epsilon$ is set to zero as an example.  The dashed line 
represents the case without a bounce. The points where the transfer 
matrices are computed in our model are  also indicated.}
\label{meff}
\end{figure}

A matching should be performed between the three considered phases. 
It can be done, as displayed in Fig.~\ref{meff}, with transition matrices defined as follows:
\begin{equation}
{\bf M} :=  \left[\begin{array}{cc}  f_k(\eta) & f^*_k(\eta)\\ \partial_{\eta} f_k(\eta) & \partial_{\eta} f^*_k(\eta) \end{array}  \right],
\end{equation}
where the Wronskian condition implies
\begin{equation}
W(f_k(\eta),f^*_k(\eta)):= \det{\bf M}=i. 
\end{equation}
The inverse of the transition matrix is then given by:
\begin{equation}
{\bf M}^{-1} :=  -i\left[\begin{array}{cc}  \partial_{\eta} f^*_k(\eta) & -f^*_k(\eta)\\ -\partial_{\eta}f_k(\eta) 
&  f_k(\eta) \end{array}  \right].
\end{equation}

The three first transition matrices are:
\begin{eqnarray}
{\bf M_1} &=& \left[\begin{array}{cc}  \frac{e^{-ik(\eta_i-\Delta\eta)}}{\sqrt{2k}}  & \frac{e^{ik(\eta_i-\Delta\eta)}}{\sqrt{2k}} \\  
-i\sqrt{\frac{k}{2}} e^{-ik(\eta_i-\Delta\eta)}  &  i\sqrt{\frac{k}{2}} e^{ik(\eta_i-\Delta\eta)} \end{array}  \right],    \\
{\bf M_2} &=&  \left[\begin{array}{cc} \frac{e^{-i\Omega(\eta_i-\Delta\eta)}}{\sqrt{2\Omega}}  & 
\frac{e^{i\Omega(\eta_i-\Delta\eta)}}{\sqrt{2\Omega}} \\  
-i\sqrt{\frac{\Omega}{2}} e^{-i\Omega(\eta_i-\Delta\eta)}  &  i\sqrt{\frac{\Omega}{2}} e^{i\Omega(\eta_i-\Delta\eta)} 
\end{array}  \right],    \\
{\bf M_3} &=&  \left[\begin{array}{cc} \frac{e^{-i\Omega\eta_i}}{\sqrt{2\Omega}}  & 
\frac{e^{i\Omega\eta_i}}{\sqrt{2\Omega}} \\  
-i\sqrt{\frac{\Omega}{2}}e^{-i\Omega\eta_i}  & i\sqrt{\frac{\Omega}{2}} e^{i\Omega\eta_i} \end{array}  \right],
\end{eqnarray}
where 
\begin{equation}
\Omega = \sqrt{k^2+k^2_0}.
\end{equation}

In the last region, mode functions can be written as
\begin{equation}
f_k(\eta) =  \alpha_k g_k(\eta) +\beta_k g^*_k(\eta), \label{fdec}
\end{equation}
where 
\begin{equation}
g_k(\eta) = \sqrt{-\eta} \sqrt{\frac{\pi}{4}} e^{i\pi(2\nu+1)/4} H^{(1)}_{\nu}(-k\eta), \label{Hankel}
\end{equation}
$H_{\nu}(x)$ being a Hankel function of the first kind. The mode functions $g_k(\eta)$
correspond to another decomposition of the field $\hat{u}_{\bf k}(\eta)$ in the 
form:
\begin{equation}
\hat{u}_{\bf k}(\eta) = g_k(\eta) \hat{a}_{\bf k} +g_k^*(\eta) \hat{a}^{\dagger}_{\bf-k}. \label{decomp2}
\end{equation}
The  creation ($\hat{a}^{\dagger}_{\bf k}$) 
and annihilation ($\hat{a}_{\bf k}$) operators fulfill 
the commutation relation $[\hat{a}_{\bf k},\hat{a}^{\dagger}_{\bf q}]=\delta^{(3)}({\bf k-q})$. 
Because decompositions (\ref{decomp1}) and (\ref{decomp2}) are equivalent, based on Eq.~(\ref{fdec})
and on the Wronskian conditions for the mode functions $f_k$ and $g_k$, one obtains:
\begin{equation}
\left[\begin{array}{c} \hat{b}_{\bf k}  \\ \hat{b}^{\dagger}_{\bf- k}   \end{array}  \right] = 
\left[\begin{array}{cc} \alpha_k  & \beta_k^* \\  
\beta_k & \alpha^*_k  \end{array}  \right] \left[\begin{array}{c} \hat{a}_{\bf k}  \\ \hat{a}^{\dagger}_{\bf - k}  
\end{array}  \right], \label{transform} 
\end{equation}
which corresponds to a Bogoliubov transformation with coefficients $\alpha_k$ and $\beta_k$. 
Because of the commutation
relation of the creation and annihilation operators we have  $|\alpha_k|^2-|\beta_k|^2=1$. It is 
clear from Eq.~(\ref{transform}) that if $\beta_k\neq 0$ 
particles are created from the vacuum, just because 
$\hat{b}_{\bf k} | 0\rangle = \beta_k^* \hat{a}^{\dagger}_{\bf -k} | 0\rangle$. By matching  the 
three regions, the unknown coefficients 
$\alpha_k$ and $\beta_k$ can be determined:
\begin{eqnarray}
\left[\begin{array}{c} \alpha_k  \\ \beta_k \end{array}  \right] &=& 
{\bf M_4}^{-1}{\bf M_3}{\bf M_2}^{-1}{\bf M_1}\left[\begin{array}{c} 1 \\ 0 \end{array}  \right] \label{match} \\
 &=& {\bf M_4}^{-1} \left[\begin{array}{c} \frac{e^{i k (\Delta\eta-\eta_i)} (\Omega  \cos[\Delta \eta  \Omega ]
-i k \sin[\Delta \eta  \Omega ])}{\sqrt{2k}\Omega } \\ \frac{e^{i k (\Delta \eta-\eta_i)} 
(-i k\cos[\Delta \eta  \Omega ]-\Omega\sin[\Delta \eta  \Omega ])}{\sqrt{2k}}\end{array}\right], 
\nonumber 
\end{eqnarray}
where ${\bf M_4}$ is given by 
\begin{equation}
{\bf M_4} =  \left[\begin{array}{cc}  g_k(\eta) & g^*_k(\eta)\\ \partial_{\eta} g_k(\eta) & \partial_{\eta}
g^*_k(\eta) \end{array}  \right]_{\eta=\eta_i},
\label{M4SR}  
\end{equation}
the mode function $g_k$ being given by Eq.~(\ref{Hankel}).
In the special case corresponding to a de Sitter inflation ($\epsilon=0$ and $\nu=\frac{3}{2}$),
the mode functions given by Eq.~(\ref{Hankel}) simplify to the Bunch-Davies vacuum
\begin{equation}
\left. g_k(\eta) \right|_{\nu=\frac{3}{2}}=g^{\text{B-D}}_k(\eta) 
=\frac{e^{-ik\eta}}{\sqrt{2k}}\left(1-\frac{i}{k\eta}\right).
\end{equation}

In general, the amplitude of the mode function during inflation can be written as 
\begin{equation}
|f_k|^2=|g_k|^2|\alpha_k-\beta_k|^2 +4\Re(\alpha_k^*\beta_k g_k^*) \Re{g_k} \label{u2}.
\end{equation}
As we are interested in the spectrum at the end of inflation ($\eta \rightarrow 0^-$), the approximation  
\begin{equation}
H^{(1)}_{\nu}(x) \simeq - \frac{i}{\pi} \Gamma (\nu) \left( \frac{x}{2} \right)^{-\nu} 
\end{equation}
holds and, based on this, one can easily see that for a slow-roll inflation ($\epsilon \ll 1$):
\begin{equation}
\lim_{\eta\rightarrow 0^-} \frac{\Re{g_k(\eta)}}{\Im{g_k(\eta)}} = \mathcal{O}(\epsilon).
\end{equation}
Therefore, the leading order contribution from Eq.~(\ref{u2}) becomes
\begin{equation}
\lim_{\eta\rightarrow 0^-}|f_k|^2 =|g_k|^2|\alpha_k-\beta_k|^2. 
\end{equation}
With this approximation, the tensor power spectrum 
at the end of inflation takes the form 
\begin{equation}
\mathcal{P}_{\text{T}}(k) = \frac{16}{\pi} \left(\frac{H}{m_{\text{Pl}}} \right)^2 
\left( \frac{k}{aH}\right)^{-2\epsilon} |\alpha_k-\beta_k|^2. \label{PTanalytical}
\end{equation}
The coefficients $\alpha_k$ and $\beta_k$ are computed from Eq.~(\ref{match}). 
Since the resulting expression for $|\alpha_k-\beta_k|^2$ is very long, it is not explicitly given here. 
It exhibits the correct ultra-violet (UV) behavior, namely 
$\lim_{k\rightarrow \infty} |\alpha_k-\beta_k|^2=1$. Therefore, the UV spectrum
simplifies to 
\begin{equation}
\mathcal{P}_{\text{T}}(k\to\infty) = \frac{16}{\pi} \left(\frac{H}{m_{\text{Pl}}} \right)^2\left( \frac{k}{aH}\right)^{-2\epsilon}.
\label{Pkinf}
\end{equation}

\begin{figure}[ht!]
\centering
\includegraphics[width=8cm,angle=0]{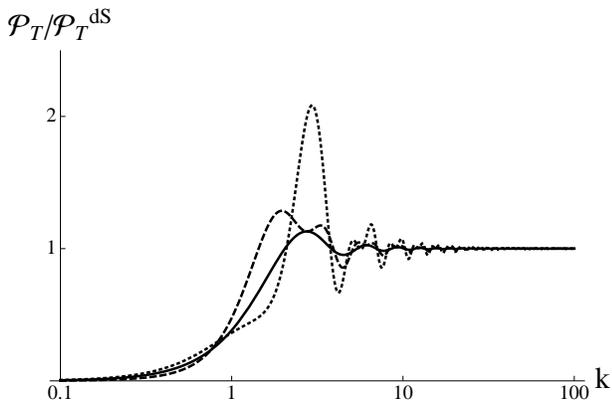}
\caption{Analytical tensor power spectra, normalized to the non-LQC-corrected spectrum, 
for three different values of $k_0$ in the  $\epsilon=0$ case. The parameters are: 
$k_0=0$ (solid line), $k_0=1.5$ (dashed line), $k_0=3$ (dotted line), 
$\eta_i=-1$, and $\Delta\eta=1$.}
\label{PT1}
\end{figure}

\begin{figure}[ht!]
\centering
\includegraphics[width=8cm,angle=0]{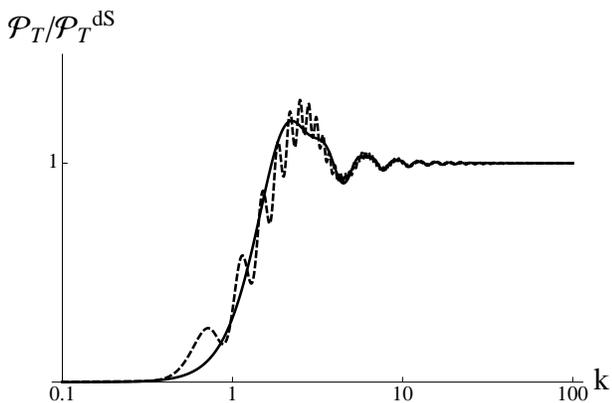}
\caption{Analytical tensor power spectra, normalized to the non-LQC-corrected spectrum,  
for  two different values of $\Delta\eta$ in the  $\epsilon=0$ case.  The parameters are: 
$\Delta\eta=0$ (solid line), $\Delta\eta=10$ (dashed line), $k_0=1$, and $\eta_i=-1$.}
\label{PT2}
\end{figure}

In Fig.~\ref{PT1}, spectra, as obtained from Eq.~(\ref{PTanalytical}), are displayed for 
different values of $k_0$ and normalized to the usual non-LQC corrected spectrum. In Fig.~\ref{PT2}, 
the width of the bounce $\Delta\eta$ is varied. In both
cases, $\epsilon$ is vanishing. 

The main features that can be drawn from those plots are the following:
\begin{itemize}
\item The power is suppressed in the infra-red (IR) regime. This is a characteristic feature associated
with the bounce.
\item The UV behavior agrees with the standard general relativistic picture.
\item Damped oscillations are superimposed with the spectrum around the "transition" momentum $k_*$ between the
suppressed regime and the standard regime.
\item The first oscillation behaves like a "bump" that can substantially exceed the UV asymptotic
value.
\item The parameter $k_0$ basically controls the amplitude of the oscillations whereas $\Delta\eta$
controls their frequency. 
\end{itemize} 

\section{Numerical investigation of the power spectrum}

To perform a more detailed analysis, we have also fully numerically solved the system of coupled
differential equations which leads to both the evolution of the modes and of the background:
\begin{eqnarray}
\frac{d^2 f_k}{dt^2} &=&-H \frac{df_k}{dt}-\left[\frac{k^2}{a^2}+\frac{\kappa}{6} \left(3p - \rho \right) \right]f_k,\\
\frac{dH}{dt} &=& \frac{1}{2} \kappa (\rho+p) \left(2
\frac{\rho}{\rho_c} - 1\right), \\
\frac{da}{dt} &=& H a, \\
\frac{d\phi}{dt} &=& \frac{\pi_{\phi}}{a^3}, \\
\frac{d\pi_{\phi}}{dt} &=& - a^3\phi, 
\end{eqnarray}
where 
\begin{equation}
\rho=\frac{\pi^2_{\phi}}{2a^6}+\frac{m^2}{2}\phi^2 \  \text{and} \  p=\frac{\pi^2_{\phi}}{2a^6}-\frac{m^2}{2}\phi^2
\end{equation}
are respectively the energy density and pressure of the scalar field whereas 
$\pi_\phi$ is the momentum.

To compute the evolution of the modes, the initial condition was assumed to be the 
Minkowski vacuum
\begin{equation}
f_k = \frac{e^{-ik\eta}}{\sqrt{2k}}.
\end{equation}
This approximation is valid for the subhorizontal modes. Therefore, in the numerical computations
we have evolved only modes that were subhorizontal at the initial time. 

\begin{figure}[ht!]
\centering
\includegraphics[width=8cm,angle=0]{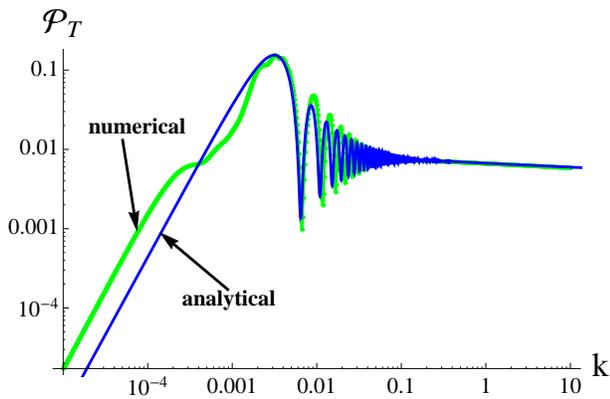}
\caption{Comparison of numerical and analytical spectra Eq.~(\ref{PTanalytical}) for $m=10^{-2}m_{\text{Pl}}$. 
In the IR region the spectra behave as $\mathcal{P}_{\text{T}} \propto k^2$ while in the UV region
they behave as 
$\mathcal{P}_{\text{T}} \propto k^{-2\epsilon}$, where $\epsilon \ll 1$ is the slow-roll parameter. 
Here: $H=0.037 m_{\text{Pl}}, \epsilon = 0.0246, k_0=0.037 m_{\text{Pl}}, \eta_i=-750,$ and $\Delta\eta=10$.}
\label{num}
\end{figure}

In Fig.~\ref{num}, the analytical spectrum  Eq.~(\ref{PTanalytical}) evaluated as explained in the 
previous section is compared with
the full numerical computation. The overall agreement is very good with slight deviations due to subtle
dynamical effects. The UV tilt associated with the slow-roll parameter is perfectly recovered. 
The values of parameters $H$, $k_0$ and $\epsilon$ were determined from the evolution of 
the background. In turn, the parameters $\eta_i$ and $\Delta\eta$ were fixed to fit the numerical data.

The mass of the scalar field is, of course, the key physical parameter of this model. 
The canonically chosen value (around $10^{-6}m_{\text{Pl}}$) may not be especially 
meaningful in this approach as the standard requirements of inflation are substantially 
modified by the specific history of the Hubble radius. This value is nonetheless 
still the mostly preferred one. 

\begin{figure}[ht!]
\centering
\includegraphics[width=8cm,angle=0]{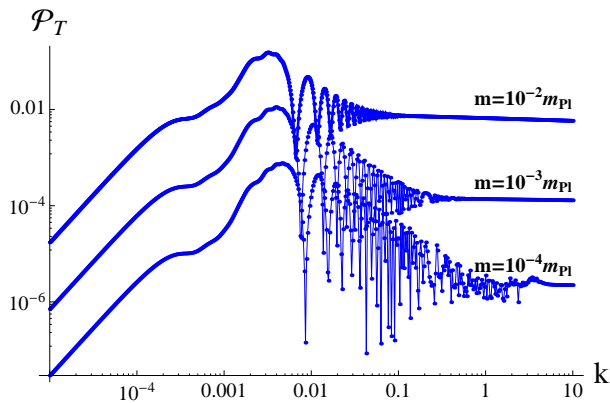}
\caption{Numerically computed power spectra for $m=10^{-4},10^{-3},10^{-2}$
$m_{\text{Pl}}$ (from bottom to top in the UV range).}
\label{varm}
\end{figure}

In Fig.~\ref{varm}, the spectra computed for three different mass values are displayed.
As expected, the UV value of the spectrum scales as $m^2$, since during inflation 
$\mathcal{P}_{\text{T}} \sim H^2 \sim m^2$.
It is also clear that the region of oscillations becomes broader while lowering the value of $m$. 

\begin{figure}[ht!]
\centering
\includegraphics[width=8cm,angle=0]{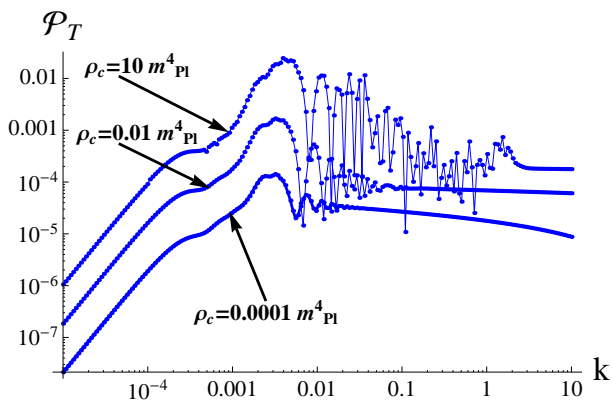}
\caption{Numerically computed power spectra for $\rho_{\text{c}}=10^{-4},10^{-2},10$
$m^4_{\text{Pl}}$ (from bottom to top in the UV range) with $m=10^{-3}m_{\text{Pl}}$.}
\label{varrho}
\end{figure}

In Fig.~\ref{varrho}, we show how the spectrum is modified by different choices of 
$\rho_{\text{c}}$.
It is clear that increasing $\rho_{\text{c}}$ leads to an amplification of the spectrum. 
{The dependence is however not very strong. As it was shown in Section \ref{SecII}, 
the increase of $\rho_{\text{c}}$ leads to an increase of the field displacement 
$\phi_{\text{max}}$. This dependence was shown to be rather weak. Since 
$\mathcal{P}_{\text{T}}\sim H^2 \sim m^2\phi^2$, the increase of  
$\phi$ due to the dependence upon $\rho_{\text{c}}$ will result in an amplification of the power
spectrum. This is in agreement with  the numerical results. From Fig. \ref{varrho}, it can also 
be noticed that increasing $\rho_{\text{c}}$ amplifies the oscillatory structure.

The numerical investigations performed for this work have shown that the quantity $R$ defined as
\begin{equation}
R := \frac{\mathcal{P}_{\text{T}}(k=k_*)}{\mathcal{P}^{\text{standard}}_{\text{T}}(k=k_*) }, 
\label{Rdef}
\end{equation}
basically evolves as
\begin{equation}
R \simeq \left(\frac{m_{\text{Pl}}}{m} \right)^{0.64},  \label{R}
\end{equation}
where $k_*$ is the position of the highest peak in the power spectrum and
$\mathcal{P}^{\text{standard}}_{\text{T}}(k)$ is a standard inflationary 
power spectrum [see e.g. Eq.~(\ref{Pkinf})] which
overlaps with $\mathcal{P}_{\text{T}}(k)$ for $k\rightarrow \infty$.
The function (\ref{R}) was obtained by fitting the numerical data in the mass range 
$m=5\cdot10^{-5} m_{\text{Pl}} \ ... \ 10^{-1} m_{\text{Pl}}$. Because of numerical instabilities,
it was not possible to perform computations for lower values of the
inflaton mass.  The numerically obtained values of $R$ together with the 
approximation given by Eq.~(\ref{R}) are given in
Fig.~\ref{ratio}. 
\begin{figure}[ht!]
\centering
\includegraphics[width=8cm,angle=0]{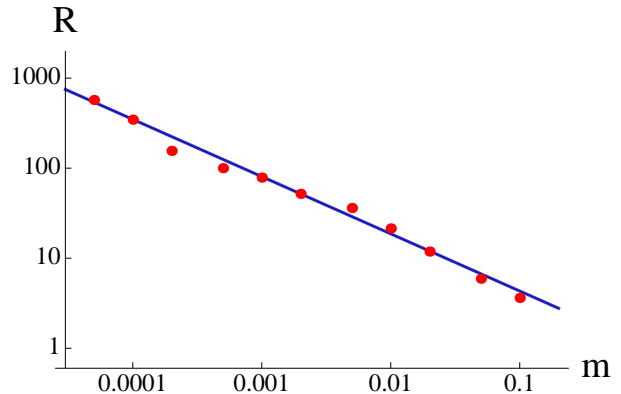} 
\caption{Ratio defined by Eq.~(\ref{Rdef}) as a function of inflaton mass in Planck units. 
Dots are values obtained from the numerical computations. The straight line
is the fit given by Eq.~(\ref{R}).}
\label{ratio}
\end{figure}
This parametrization is useful for phenomenological purposes. Interestingly, $R$ can
become very high for low values of the mass of the field. This partially compensates for
the lower overall normalization of the spectrum and can become a very specific feature 
of the model. In particular, for the 
mass  $m \approx10^{-6}m_{\text{Pl}}$ (which is the value preferred by some estimations), 
extrapolating the relation (\ref{R}) leads to $R\approx 8000$. If the relation still holds in this range,
the effect is very significant, and could have important observational consequences.
 
Finally, to make basic studies easier, we performed a rough parametrization of the full spectrum:
\begin{equation}
\mathcal{P}_{\text{T}}= \frac{16}{\pi} \left(\frac{H}{m_{\text{Pl}}} \right)^2 
\frac{\left( \frac{k}{aH}\right)^{-2\epsilon} }{1+(k_*/k)^2} 
\left[1+\frac{4R-2}{1+(k/k_*)^2} \right],  \label{eff1}
\end{equation}
leading to
\begin{equation}
\mathcal{P}^{\text{dS}}_{\text{T}}= \frac{16}{\pi} \left(\frac{H}{m_{\text{Pl}}} \right)^2 \frac{1}{1+(k_*/k)^2} 
\left[1+\frac{4R-2}{1+(k/k_*)^2} \right],  \label{eff2}
\end{equation}
in the specific case of de Sitter inflation. In both cases, the classical behavior is recovered in the
limit $k \rightarrow \infty$. The point for introducing the $R$ factor the way it was done becomes 
clear when calculating the
value of the spectra at $k=k_*$. For a modified de Sitter spectrum [Eq.~(\ref{eff2})], we get
\begin{equation}
\mathcal{P}^{\text{dS}}_{\text{T}}(k=k_*) = R  \frac{16}{\pi} \left(\frac{H}{m_{\text{Pl}}} \right)^2.
\end{equation}  
Thanks to the relation (\ref{R}), the number of the free parameters can be decreased in a 
phenomenological analysis.

As shown on Fig.~\ref{param}, this formula correctly reproduces the main features, namely the IR power
suppression, the bump and the UV limit. Oscillations are missed but due to momentum integration there is
little hope that they can observationally be seen on a cosmological microwave background (CMB) spectrum.
\begin{figure}[ht!]
\centering
\includegraphics[width=8cm,angle=0]{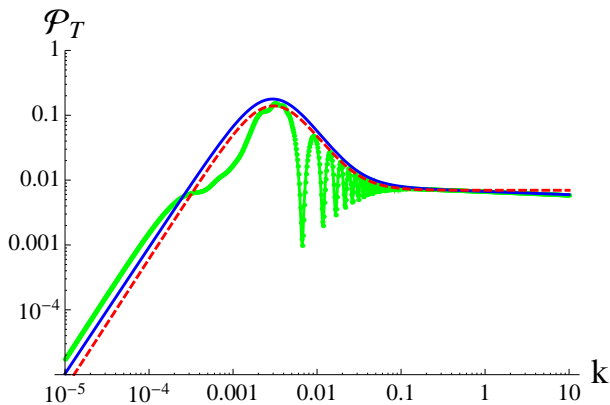}
\caption{Comparison of the numerical spectrum for $m=10^{-2}m_{\text{Pl}}$ with formulas (\ref{eff1}) and (\ref{eff2}).
The solid (blue) line corresponds to (\ref{eff1})  while the dashed (red) line  corresponds to (\ref{eff2}). }
\label{param}
\end{figure}

To conclude this section, we have schematically represented the evolution of the Hubble
radius ($R_{\text{H}}:=1/|H|$), together with the  physical modes, in Fig.~\ref{hubble}. This helps
to understand the shape of the obtained spectra.

\begin{figure}[ht!]
\centering
\includegraphics[width=8cm,angle=0]{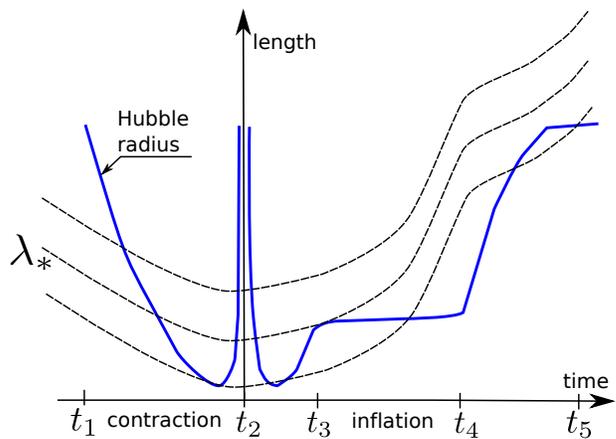}
\caption{Schematic picture of evolution of the Hubble radius (solid line) and of 
the different length scales (dashed lines) for the considered model of
the universe. Different times are distinguished: 
$t_1-$time when the initial conditions are set;  
$t_2-$bounce ($H=0$);  
$t_3-$beginning of inflation;  
$t_4-$end of inflation; 
$t_5-$present epoch of dark energy domination.}
\label{hubble}
\end{figure}

We consider the modes that are initially (at time $t_1$) shorter than the Hubble radius.
For those modes, the normalized solution is given by the Minkowski vacuum  $f_k=e^{-ik\eta}/\sqrt{2k}$.
Therefore, the initial power spectrum takes the form $\mathcal{P}_{\text{T}} \sim k^3 |f_k|^2 \sim k^2$. 
Starting from the largest scales, the modes cross the Hubble radius.  This is possible since the Hubble
radius undergoes contraction faster than any particular length scale. While crossing the 
horizon, the shape of the spectrum becomes \emph{frozen} in the initial $\mathcal{P}_{\text{T}}\sim k^2$
form. Then, the modes evolve through the bounce (at time $t_2$) until the beginning of inflation (at time $t_3$).
The main consequence of the transition of modes through the bounce is the appearance of
additional oscillations in the spectrum.
This issue was studied in detail in Ref.~\cite{Mielczarek:2009vi}, where the spectrum at  time $t_3$
was calculated for the symmetric bounce model. After the bounce, modes with wavelengths
shorter than $\lambda_*$ start to reenter the Hubble radius. The superhorizon modes 
$\lambda > \lambda_*$ ($k<k_*$) 
hold the $k^2$ spectrum,  with however some oscillatory features due to the  bounce.  Modes 
with $\lambda < \lambda_*$ ($k>k_*$)  cross the horizon again during the phase of inflation. 
For them, the spectrum agrees with the standard slow-roll inflation spectrum 
$\mathcal{P}_{\text{T}}\sim k^{-2\epsilon}$ where
$\epsilon \ll 1$. The small tilt is due to a slow increase of the Hubble radius. Contributions from
different modes are then slightly different. At the end of inflation (at time $t_4$) the spectrum is therefore 
suppressed 
($\mathcal{P}_{\text{T}}\sim k^2$) for $k<k_*$ and exhibits the inflationary shape
($\mathcal{P}_{\text{T}}\sim k^{-2\epsilon}$)
for $k>k_*$. The spectrum is also modified by the oscillations due to the bounce. This corresponds 
to the computations of this paper. The particular mode with wavelength $\lambda_*$ 
(wave number $k_*$) should be studied in more detail.  The size of this mode 
overlaps with the size of the Hubble radius at the beginning of 
inflation: $k_* \simeq a(t_3) H(t_3)$. The physical length $\lambda_*$ at the scale factor $a(t)$ is therefore 
equal to  $\lambda_*(t) \simeq a(t)/[a(t_3)H(t_3)]$. This scale grows with the cosmic expansion 
and it is crucial, from the observational point of view, to determine its present size (at time $t_5$). The case drawn 
in Fig.~\ref{hubble} corresponds to a present size of $\lambda_*$ greater than the size of the horizon.  This 
is indeed rather unlikely that the present value of $\lambda_*$ is below the size of horizon just because the 
spectrum of scalar perturbations should then exhibit deviations from the nearly scale invariant inflationary 
prediction. Up to now, there is no observational evidence for such deviations. A remaining possibility would 
however be that the (slight) observed lack of power in the 
CMB spectrum of anisotropies could be due to the effects of the bounce. 
However, the present size of  $\lambda_*$ would then be comparable with the size of horizon. 
This leads to the question: why should those two scales overlap right now? 
This is rather unnatural, and would lead to a new coincidence problem. However, as it was 
estimated in Ref.~\cite{Mielczarek:2009zw}, these two scales can indeed overlap in the 
standard inflationary scenario for quite natural values of the parameters. 
There is therefore a glimpse of hope that the scale $\lambda_*$ is at least not to much bigger than the size of horizon.
This could allow us to see some UV features due to the bounce as the oscillations also affect sightly
the inflationary part of the spectrum. These are however secondary effects and it is not clear whether
they were not smoothed away during the radiation domination era.  Moreover, in the region where those 
effects could be expected, errors due to the \emph{cosmic variance} become significant. This is an unavoidable 
observational limitation which cannot be bettered, even by the improvement of resolution of the 
future CMB experiments.

Another limitation in studying the effects of LQC comes from the fact that the derived 
modifications can also appear in other bouncing cosmologies. In particular, within the model of 
quintom bounce, the discussed effects of suppression and oscillations were also pointed out 
\cite{Cai:2008ed,Cai:2008qb}. The amplitude of tensor perturbations at the peak was however 
not predicted to be as high as in LQC. An additional amplification on the very large 
scales was also predicted in the quintom model. Despite these differences, at the observationally 
accessible low scales, the effects due to the LQC bounce and the quintom bounce are mostly
indistinguishable. Therefore, complementary observational methods have to be proposed to 
distinguish between such models. A possible distinction could be given e.g. from the analysis 
of non-Gaussianity production within LQC.

\section{Conclusions}

This study establishes the full background dynamics in bouncing models with
holonomy corrections. Although this has already been claimed before, we confirm
that due to the sudden change of sign of the Hubble parameter, inflation is
nearly unavoidable. In this paper, we have considered a particular model of inflation 
where the content of the universe is dominated by a massive scalar field. We have 
investigated the details by both analytical and numerical studies the primordial
power spectrum of gravitational waves. It exhibits several characteristic features, 
namely a $\mathcal{P}_{\text{T}}\propto k^2$ 
IR power suppression, oscillations, and a bump at $k_*$. In the UV  regime, the 
standard inflationary spectrum $\mathcal{P}_{\text{T}}\propto k^{-2\epsilon}$ is recovered.
The primordial tensor power spectrum transforms into $B$-type CMB polarization. The 
performed investigations therefore open the window for observational tests of the model, 
in particular through the amplification which occurs while approaching $k \rightarrow k_*$. 
The observed structures correspond to the UV region in the spectrum. If the present 
scale $\lambda_* \sim 1/k_*$ is not much larger than the size of horizon, then the effects
of the bounce should be, in principle, observable.  In particular, one should expect 
amplification, rather than suppression of the $B$-type polarization spectrum at the low multipoles. 
The suppression for $k<k_*$ becomes dominant at the much larger scale, probably
far above the horizon.  While the $B$-type polarization has not been detected yet, 
there are huge efforts in this direction. 
Experiments such as PLANCK \cite{:2006uk}, BICEP \cite{Chiang:2009xsa} or
QUIET \cite{Samtleben:2008rb} are (partly) devoted to the search of the $B$ mode.
Even with present observational constraints, one can already exclude some evolutionary scenarios
and possible values of the parameters, in particular the inflaton mass $m$ and position of 
the bump $k_*$ in the spectrum. We address this interesting issue elsewhere \cite{dream_team}.
There are also still several points to study around this model:
\begin{itemize}
\item How is the scenario modified when quantum backreaction is taken into account (in particular
when the potential energy of the field in not negligible at the bounce)?
\item How is the power spectrum modified by inverse-volume terms in this framework? Although the
background dynamics should not be fundamentally altered, the spectrum could be significantly modified.
\item How do those results compare with models dealing with {\it classical} bounces (see, {\it e.g.},
\cite{peter})? If the IR power suppression is probably a generic feature of bounces, the detailed
features are model-dependent.  
\end{itemize}

Together with the known success of LQC (The singularity resolution, the correct low-energy behavior,
etc.), the facts that 1) inflation naturally occurs and 2) observational features can be expected from
the model, are strong cases for loop cosmology. Those two points are the main results of this paper.

\acknowledgments

This work was supported by the Hublot - Gen\`eve company. 
JM has been supported by the fellowship from the Florentyna Kogutowska Fund, 
from the Astrophysics Poland-France Fund and by Polish Ministry of Science 
and Higher Education grant N N203 386437. JG acknowledges financial support 
from the Groupement d'Int\'er\^et Scientifique (GIS) 'consortium Physique des 2 Infinis~(P2I)'.

\end{document}